# High Precision Digitizer for High Luminosity LHC

Miguel Bastos, Nikolai Beev, Michele Martino

High Precision Measurements section, Electrical Power Converters group, CERN

**Abstract:**

*One of the key elements of the HL-LHC project is the replacement of the magnets that focus the beams near the interaction points of ATLAS and CMS. The new magnets also call for higher precision powering, which is strongly dependent on the performance of the electric current measurement chain. We present a new metrology-grade digitizer developed at CERN, which is part of that high-precision measurement system.*

This article presents the new digitizer to be employed in the Inner Triplet and separation/recombination dipole magnet power converters of the HL-LHC project. This Analog-to-Digital Converter (ADC), named HPM7177, was designed at the High Precision Measurements section, Electrical Power Converters group, Technology department at CERN, and first tested in 2019.

The main goal of the High Luminosity LHC (HL–LHC) project [1] is to increase the instantaneous luminosity of the LHC beam. For that purpose, the project foresees the replacement of several magnets in the LHC. Among the most important are the Inner Triplets, located on each side of the interaction points of the ATLAS and CMS experiments. The Inner Triplets contribute to increasing luminosity by reducing the beam size at the interaction point. This improvement requires a fine tuning of the beam parameters, which translates into unprecedented performance requirements for the magnetic field, and consequently for the electric current that generates it [2].

Electrical power converters are used to power the magnets in particle accelerators. They are commonly employed as controlled current sources. At CERN, power converters use high precision current feedback loops implemented in the digital domain to deliver the magnet currents. As a consequence, both the reference current and the measured current need to be provided in the digital domain. The reference current is sent digitally by the control room. However, the current in the magnets is measured in the analog world, using a Direct-Current Current Transformer (DCCT) and therefore needs to be converted into a digital code. A high precision ADC is used for that purpose.

The demand for high precision of the electrical current delivered to the magnets is described in detail in [2]. The requirements for the power converters are unprecedented in terms of current stability, noise, and repeatability. Since power converter performance depends greatly on the quality of the measurement used for the feedback, the DCCT and the ADC are crucial for delivering the required precision. The performance requirements for the HL-LHC Inner Triplets power converter ADCs are described in [3]. Short-term stability (1 mHz < f < 100 mHz) of 0.05 ppm (parts per million) rms, 12 h stability of 0.2 ppm p-p (in isothermal conditions) and linearity of 1 ppm, are just a few of the challenging requirements imposed on the ADCs.

The HPM7177 digitizer is an entire standalone measurement system. Its core element is a commercial high-resolution ADC integrated circuit, selected after an extensive market survey [4] and test campaign. The digitizer employs precision circuits for the scaling of the analog signals. Digital logic functionality is implemented in a field-programmable gate array (FPGA), which takes care of the ADC chip initialization and readout, the communication and synchronization protocol, as well as the built-in calibration and self-test features of the system.

Stable measurements on the timescale of a typical LHC cycle encounter numerous challenges. Electronic components exhibit different kinds of noise, including the omnipresent 1/f or "flicker" type with spectral density rising at low frequencies. External influences such as temperature variations and electromagnetic interference (EMI) also affect the measurements. All these factors are addressed by different aspects of the HPM7177 design. To minimize low-frequency electronic noise, the voltage scaling circuits employ bulk metal foil resistors and auto-zero operational amplifiers. Ultimately, the performance on longer timescales is limited by the voltage reference, which is the best one presently available on the market. The digitizer has very low temperature-dependent drift on the order of tens of ppb (parts per billion) per degree Celsius, achieved by active temperature stabilization of the sub-module that contains all precision circuits. On the system level, multiple measures are taken to ensure EMI immunity, since the ADCs often have to operate in a potentially noisy environment near power converters.

A characterization campaign carried out using reference equipment from the CERN standards laboratory proved that HPM7177 meets the most challenging requirements for the HL-LHC project. To gain more confidence and knowledge of the device, we have planned to collaborate with the German national institute of metrology (PTB – Braunschweig) to characterize the digitizer using their 10 V Programmable Josephson Voltage Synthesizer [5]. This system generates test voltages with ultimate stability using the Josephson effect – a quantum phenomenon that links frequency to voltage and currently represents the Volt in the International System of Units (SI).

The full design documentation of HPM7177 is available for free under the CERN Open Hardware License [6].

[1] G. Apollinari, I. Béjar Alonso, O. Brüning, P. Fessia, M. Lamont, L. Rossi, and L. Tavian, "HL-LHC Technical Design Report," Tech. Rep. EDMS n. 1723851 v.0.71, CERN, Geneva, 2016.
https://edms.cern.ch/document/1723851/0.71

[2] Update of beam dynamics requirements for HL-LHC electrical circuits. CERN-ACC-2019-0030. Gamba, Davide (CERN) ; Arduini, Gianluigi (CERN) ; Cerqueira Bastos, Miguel (CERN) ; Coello De Portugal - Martinez Vazquez, Jaime Maria (Universitat Politecnica Catalunya (ES)) ; De Maria, Riccardo (CERN) ; Giovannozzi, Massimo (CERN) ; Martino, Michele (CERN) ; Tomas Garcia, Rogelio (CERN)
https://cds.cern.ch/record/2656907?ln=en

[3] HL-LHC Power Converter, ADC and DCCT Requirements; Miguel Cerqueira Bastos, CERN EDMS 2048827
https://edms.cern.ch/document/2048827/2

[4] N. Beev. Analog-to-digital conversion beyond 20 bits. Proceedings of I2MTC-2018, Houston, TX (2018)
https://www.researchgate.net/publication/325285614_Analog-to-digital_conversion_beyond_20_bits_Applications_architectures_state_of_the_art_limitations_and_future_prospects

[5] Josephson Technology at PTB-Braunschweig
https://www.ptb.de/cms/en/ptb/fachabteilungen/abt2/fb-24/ag-243/forschung-243.html

[6] HPM7177 Open Hardware Repository wiki page
https://ohwr.org/project/opt-adc-10k-32b-1cha/wikis/

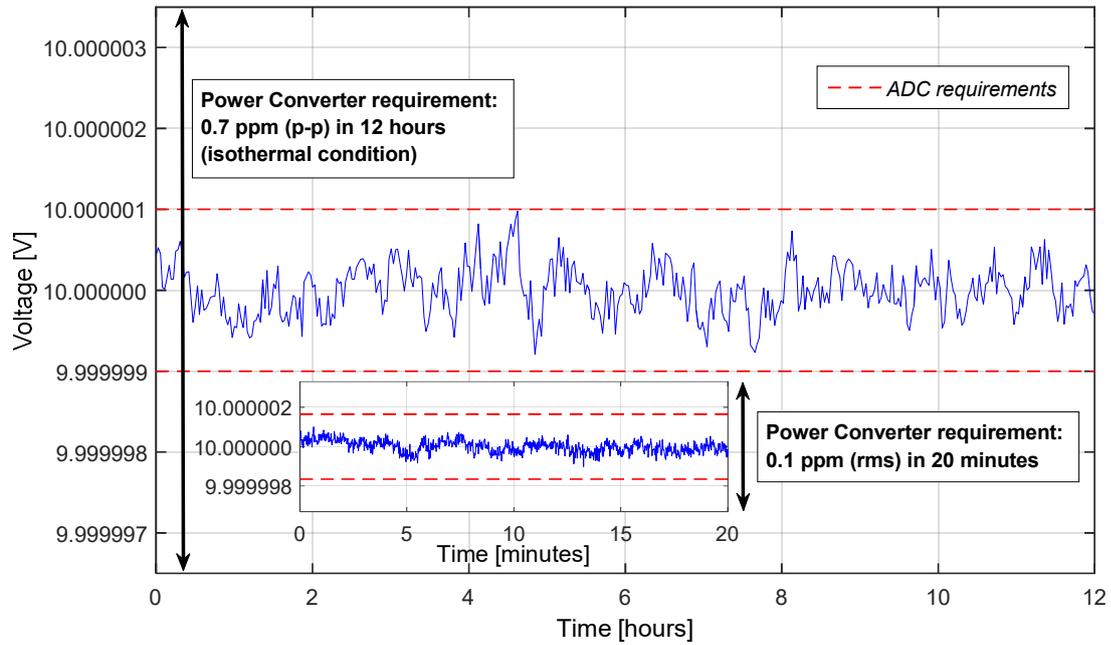

*Typical measurements at the nominal digitizer full scale of 10 V using a portable voltage standard. The main plot shows a 12-hour record, while the inset is a zoom-in to a 20-minute section of it. HL-LHC requirements are indicated by double arrows (power converter) and dashed lines (ADC).*

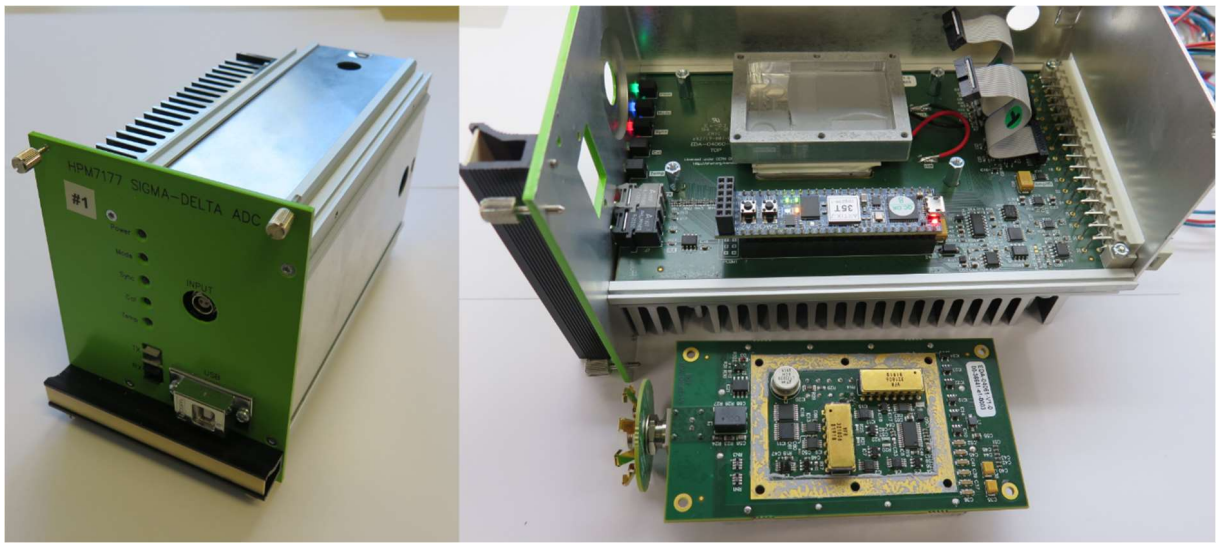